\documentclass[aps,prb,onecolumn,a4paper,10pt,showpacs,citeautoscript,notitlepage,preprintnumbers]{revtex4-1}

\usepackage{graphicx,amsfonts,epsfig,latexsym,amscd,amsmath,theorem,mathrsfs,color}
\usepackage[colorlinks=true, a4paper=true, pdfstartview=FitV,linkcolor=blue, citecolor=blue, urlcolor=blue]{hyperref}

\def\comment#1{}

\newcommand{\beg}{\begin{eqnarray}}
\newcommand{\eee}{\end{eqnarray}}

\def\cm#1{}

\newcommand{\hh}{{\cal H}}
\newcommand{\eps}{\epsilon}

\newcommand{\f}{\frac}

\newcommand{\be}{\begin{equation}}
\newcommand{\ee}{\end{equation}}
\newcommand{\ba}{\begin{eqnarray}}
\newcommand{\ea}{\end{eqnarray}}
\newcommand{\beq}{\begin{equation}}
\newcommand{\eeq}{\end{equation}}
\newcommand{\bea}{\begin{eqnarray}}
\newcommand{\eea}{\end{eqnarray}}
\newcommand{\bastar}{\begin{eqnarray*}}
\newcommand{\eastar}{\end{eqnarray*}}

\newcommand{\cd}{\partial}

\newcommand{\ignore}[1]{}

\newcommand{\ie}{{\it i.e.~}}
\newcommand{\resp}{\emph{resp.~}}

\begin{document}
\title{Type-1.5 superconductivity in muliband and other multicomponent systems}

\author{
E. Babaev${}^{1,2}$,  M. Silaev${}^{1,3}$ }
\affiliation{
${}^1$Department of Theoretical Physics, The Royal Institute of Technology, Stockholm, SE-10691 Sweden\\
${}^2$ Department of Physics, University of Massachusetts Amherst, MA 01003 USA\\
${}^3$ Institute for Physics of Microstructures RAS, 603950 Nizhny Novgorod, Russia.\\
}

\begin{abstract}

Usual
superconductors are classified into two categories as follows: type-1 when the ratio of 
the magnetic field penetration length ($\lambda$) to coherence length ($\xi$) $ \kappa=\lambda/\xi <1/\sqrt{2}$ and type-2 when $\kappa >1/\sqrt{2}$.
The boundary case $\kappa =1/\sqrt{2}$ is also considered to be a special situation, frequently termed as ``Bogomolnyi limit".
Here we discuss multicomponent systems which
can possess three or more fundamental length scales and  allow a
separate  superconducting state, which was recently termed
``type-1.5". In that state a system has the following hierarchy of coherence and penetration lengths $\xi_1<\sqrt{2}\lambda<\xi_2$. We also briefly overview the works on
single-component regime $\kappa \approx 1/\sqrt{2}$ and comment on recent discussion by Brandt and Das in the proceedings of the previous conference in this series.
{\sl Prepared for the proceedings of International Conference on Superconductivity and Magnetism 2012}

\end{abstract}
\maketitle
\section{Introduction}
\subsubsection{Type-1, type-2 and type-1.5 superconductivity.}
The fundamental classification of superconductors is based on the
 their description in terms of the classical field theory: the Ginzburg-Landau (GL) model,
and the fundamental length scales which it yields:  the coherence
length $\xi$ and the  magnetic field penetration length $\lambda$.
 There  type-1 and type-2 regimes can be distinguished in many ways,
through magnetic response, properties and stability of topological defects,
orders of the phase transitions etc.

Type-1 superconductors expel weak magnetic fields, by generating surface currents. In stronger fields
finite-size samples  form non-universal configuration of
 macroscopic normal domains with magnetic flux \cite{GL,deGennes,prozorov}.
The supercurrent in that case flows both on the surface and near the boundaries of these domains.
The response of type-2 superconductors is
   different \cite{abrikosov}; below some critical value $H_{c1}$, the field is expelled. Above this value a superconductor forms a lattice
or a liquid of vortices which carry magnetic flux through the
system. Since vortices {are formed by the supercurrent} it also
implies that type-2 superconductor in external field allow
 current configuration in its bulk (in contrast to only surface currents in type-1 case).
 Only at a higher second critical value, $H_{c2}$ superconductivity is destroyed.

The special ``zero measure'' boundary case where $\kappa$ has a
critical value exactly at the type-1/type-2 boundary is
legitimately  distinguished as a separate case (frequently called
``Bogomolnyi regime"). In the most common GL model
parameterization it corresponds to
 $\kappa = 1/\sqrt{2}$.
In that
case stable vortex excitations exist but they
do not interact \cite{kramer,bogomolny} in the Ginzburg-Landau theory.
This regime is a subject of quite broad interest also beyond condensed matter context \cite{mansut}.

It was recently shown  that in {\it multicomponent} systems
systems (in particular in multiband superconductors)
there is a regime which falls outside the type-1/type-2 dichotomy
 \cite{bs1,bcs,bcs2,moshchalkov,garaud,silaev,silaev2,moshchalkov2,review,daonew,3bands,sro,vargunin},
 see also recent related works \cite{recent,recent2,recent22,recent2a,shanenko}
In that case the system possesses two or more coherence lengths
such that (in the most general N-component case)
 $\xi_1,..,\xi_k<\sqrt{2}\lambda<\xi_{k+1},...,\xi_N$ and there are thermodynamically stable vortices with
long-range attractive, short range repulsive interaction, as a consequence of this hierarchy of length scales. Owing to its multicomponent
nature there are two kinds of superflows in the system  and they
can have {\it coexisting} type-1 and type-2 tendencies, and do not
fall exclusively under the definitions of either
type-1 or type-2 cases.

\subsubsection{Bogomolnyi regime $\kappa = 1/\sqrt{2}$}

Let us start by making some remarks about a  particular 
limit of {\it single component} Ginzburg-Landau theory: the Bogomolnyi regime
$\kappa = 1/\sqrt{2}$.   It is a property of Ginzburg-Landau model
where, at $\kappa = 1/\sqrt{2}$, the core-core attractive
interaction between vortices exactly cancels the current-current
repulsive interaction at all distances as  was first
discussed by Kramer and later in more detail by Bogomolnyi \cite{kramer,bogomolny}. For a review of current
studies of that regime in the Ginzbug-Landau theory see \cite{mansut}.
 However indeed in a realistic system even in the limit
 $\kappa = 1/\sqrt{2}$, there will be always
  leftover inter-vortex interactions,
(appearing beyond the GL field theoretic description),
 form underlying
microscopic physics. The form of that interaction potential is
determined  not by the fundamental length scales of the GL theory
but by  non-universal microscopic physics and in general it can
indeed be non-monotonic.

 The idea of searching for non-monotonic
intervortex potential from microscopic corrections in the
Bogomolnyi regime  $\kappa \approx 1/\sqrt{2}$ was suggested by Eilenberger and Buttner \cite{Eilenberger},
and later by Halbritter,
Dichtel   \cite{Eilenberger} \cite{Halbritter, Dichtel}.
Unfortunately, as was later discussed 
by b Leung and
Jacobs these early works were { based
on uncontrollable approximations}\cite{LeungJacobs,Leung}.
Nonetheless the importance of
Eilenberger's work cannot be denied as it appears to be the first
to raise the question if the cancellation of intervortex interaction
forces in the Ginzburg-Landau model with $\kappa=1/\sqrt{2}$ can give raise to
the extretemely weak but qualitatively interesting microscopic-physics dominated intervortex forces. 
It should be noted that all of the above works
do not discuss universal physics, but
are focused on specific microscopic physics
of a weakly coupled BCS superconductor.

\section{Type-1.5 superconductors}
In this paper we focus on what kind of new physics
can arise in {\it multicomponent} superconductors.
We argue  that type-1/type-2 dichotomy breaks down in these systems
despite various couplings between the components.

 The Ginzburg-Landau free energy functional for multicomponent system
 has the form

\begin{equation}
F=\frac{1}{2}\sum_i (D\psi_i)(D\psi_i)^* + V(|\psi_i|)
+\frac{1}{2}(\nabla\times {\bf A})^2
\label{gl0}
\end{equation}
Here $\psi_i$ are complex superconducting components,
$D=\nabla + ie {\bf A}$, and $\psi_a=|\psi_a|e^{i\theta_a}$,
$a=1,2$,  and $V(|\psi_i|)$ stands for effective potential.
Depending on symmetry of the system there can also
be present mixed (with respect to components $\psi_i$)  gradient terms
(for a more detailed review see \cite{bcs2,review}).

The main situations where multiple superconducting components arise are
 (i) multiband superconductors \cite{suhl,mgb,xi,gurevich1,asker,gurevich2,zhitomirsky,golubev,agterberg,iron,li} (where $\psi_i$ represent
condensates belonging to different bands), (ii) mixtures of
independently conserved condensates such as the projected
superconductivity in metallic hydrogen and hydrogen rich alloys
\cite{ashcroft,Nature,Nature2,sublattice,herland}, (where $\psi_i$
represent electronic and protonic Cooper pairs or deuteronic
condensate) or models of nuclear superconductors in neutron stars
interior  \cite{ns} (where $\psi_i$ represent protonic and
$\Sigma^-$ hyperonic condensates) and (iii) superconductors { with
nontrivial} pairing symmetries. The principal difference between
the  cases (i) and (ii) is the absence of the intercomponent
Josephson coupling in case of system like metallic hydrogen (ii)
because there the condensates are independently conserved. Thus
the symmetry is $U(1)\times U(1)$ or higher. In the case (i)
multiple superconducting components originate from Cooper pairing
in different bands. Because condensates in different bands are not
independently conserved there is a rather generic presence of
intercomponent Josephson coupling $\frac{\eta}{2}(\psi_1\psi_2^*
+\psi_2\psi_1^*)$ in that case.

\subsection{Type-1.5 superconductivity}

The possibility of a new type of superconductivity, distinct from
the type-1 and type-2 
\cite{bs1,bcs,bcs2,moshchalkov,silaev,silaev2,garaud} comes from the
following considerations. As discussed in
\cite{bs1,bcs,bcs2,moshchalkov,silaev,silaev2,garaud}, the two-component
models in general are characterized by three fundamental length
scales: magnetic field penetration length $\lambda$ and two
coherence lengths $\xi_1,\xi_2$ which renders the model impossible to
parameterize in terms of a single dimensionless parameter $\kappa$
and thus the type-1/type-2 dichotomy is not sufficient for
classification. Rather, in a wide range of parameters, as a
consequence of the existence of three fundamental length scales,
there is a separate superconducting regime with
$\xi_1/\sqrt{2}<\lambda<\xi_2/\sqrt{2}$. In that regime as a
consequence of coexisting type-1 and type-2 tendencies the following
situation is possible: vortices can  have long-range attractive
(due to ``outer cores" overlap) and  short-range repulsive interaction
(driven by current-current and electromagnetic interaction) and
form vortex clusters immersed in domains of two-component Meissner
state \cite{bs1,bcs}. It should be noted that the non-monotonic
intervortex interaction is one of the necessary properties of type-1.5 regime
but is not a defining property (more details of it is given below, see also \cite{review}).
Here we summarize the basic properties of type-1, type-2 and type-1.5 regimes in
the table \ref{table1} \cite{bcs2}.
 Recent experimental works
\cite{moshchalkov,moshchalkov2} proposed that this state is
realized in
 the two-band material MgB$_2$. In Ref.  \cite{moshchalkov}  this regime
was termed ``type-1.5'' superconductivity by Moshchalkov
et al. These works resulted in increasing interest in the subject
\cite{recent,daonew,recent2,recent2a,shanenko}.
Recently a counterpart of type-1.5 regime was  discussed in
context of quantum Hall effect \cite{qhe}.

If the vortices form clusters one cannot use the usual one-dimensional
argument concerning the energy of superconductor-to-normal state boundary
to classify the magnetic response of the system.
First of all, the energy per vortex in such a case
depends on whether a vortex is placed in a cluster or not.
Formation of a single isolated vortex might be energetically
unfavorable, while formation of vortex clusters
is favorable, because in a cluster where vortices are placed
in a minimum of the interaction potential, the energy
per flux quantum is smaller than that for an isolated vortex.
Also 
 the boundary conditions for magnetic field
do not involve having $H=0$ at ($x\rightarrow  -\infty$) and  thermodynamical critical field values $H=H_{ct}$
at ($x \rightarrow    \infty$) .
Furthermore, besides the energy
of a vortex in a cluster, there appears
an additional energy characteristic associated
with the boundary of a cluster but in general that
boundary energy also depends on structure and size of a cluster.
So in that case the argument of the boundary energy is not a useful characteristic since there are
there are many different solutions for different interfaces, some of which have negative
energy while others (such as the energy of a vortex cluster boundary) have positive energy.

\begin{table*}
\begin{center}
\begin{tabular}{|p{3cm}||p{4.5cm}|p{4.5cm}|p{5cm}|}
\hline
 & {\bf  single-component Type-1 } & {\bf single-component Type-2}  & {\bf multi-component Type-1.5}  \\ \hline \hline
{\bf Characteristic lengths scales} & Penetration length $\lambda$ \&   coherence length $\xi$ ($\frac{\lambda}{\xi}< \frac{1}{\sqrt{2}}$) & Penetration length $\lambda$ \& coherence length $\xi$ ($\frac{\lambda}{\xi}> \frac{1}{\sqrt{2}}$) & Two characteristic density variations length scales $\xi_1$,$\xi_2$ and penetration length $\lambda$,  the non-monotonic vortex interaction occurs in these systems in a large range of parameters when  $\xi_1<\sqrt{2}\lambda<\xi_2$
\\ \hline
 {\bf  Intervortex interaction} &   Attractive &    Repulsive   &Attractive at long range and repulsive at short range \\ \hline
{\bf  Energy of superconducting/normal state boundary} &    Positive    & Negative   & Under quite general conditions negative energy of superconductor/normal interface inside a vortex cluster but positive energy  of the vortex cluster's boundary  \\ \hline
{\bf The magnetic field required to form a vortex} &    Larger than the thermodynamical critical magnetic field  & Smaller than thermodynamical critical magnetic field & In different cases either (i) smaller than the thermodynamical critical magnetic field or (ii) larger than critical magnetic field for single vortex but smaller than critical magnetic field for a vortex cluster of a certain critical size
\\ \hline
{\bf  Phases in external magnetic field } & (i) Meissner state at low fields; (ii) Macroscopically large normal domains at larger fields.
First order phase transition between superconducting (Meissner) and normal states & (i) Meissner state at low fields, (ii) vortex lattices/liquids at larger fields.  Second order phase transitions between Meissner and vortex states and between vortex and normal states at the level of mean-field theory. & (i) Meissner state at low fields (ii) ``Semi-Meissner state":  vortex clusters coexisting with Meissner domains at intermediate fields (iii) Vortex lattices/liquids at larger fields.  Vortices form via a first order phase transition. The transition from vortex states to normal state is second order.
\\ \hline
{\bf  Energy E(N) of N-quantum axially symmetric vortex solutions} &    $\f{E(N)}{N}$  $<$ $\f{E(N-1)}{N-1}$ for all N. Vortices
collapse onto a single N-quantum mega-vortex &   $\f{E(N)}{N} >\f{E(N-1)}{N-1}$ for all N. N-quantum vortex
decays into N infinitely separated single-quantum vortices & There is a characteristic number N${}_c$ such
that $\f{E(N)}{N}$  $<$ $\f{E(N-1)}{N-1}$ for N $<$ N${}_c$, while $\f{E(N)}{N}$ $>$ $\f{E(N-1)}{N-1}$ for N
$>$ N${}_c$. N-quantum vortices decay into vortex clusters.
\\ \hline
\end{tabular}
\caption[Basic characteristics of superconductors]{Basic characteristics of bulk clean superconductors in type-1, type-2 and type-1.5 regimes. Here the most common units are used in which the value of the GL parameter
which separates type-1 and type-2 regimes  in a single-component theory is $\kappa_c=1/\sqrt{2}$.
Magnetization curves in these regimes are shown on Fig. \ref{magnetization}}
\label{table1}
\end{center}
\end{table*}


\begin{figure}
\includegraphics[width=\linewidth]{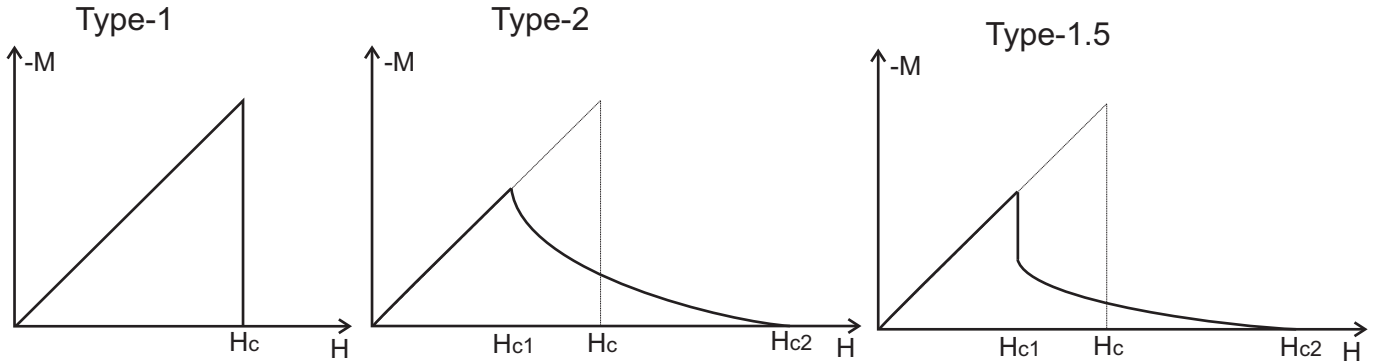}
\caption{A schematic picture of magnetization curves of type-1,
type-2 and type-1.5 superconductors.} \label{magnetization}
\end{figure}

\section{Length scales and type-1.5 regime in two-band Ginzburg-Landau Model with   arbitrary interband
interactions. }
Type-1.5 regimes with $\xi_1/\sqrt{2}<\lambda<\xi_2/\sqrt{2}$ has a clear
interpretation in $U(1)\times U(1)$ superconductors, which is the simplest example
of a superconductor which cannot be parameterized by Ginzburg-Landau parameter $\kappa$.
 Here we overview how coherence length are defined
in  Ginzburg-Landau model for two-band superconductors. These
systems have    non-zero interband interactions. Consider the most general GL
model for two-band superconductors
\begin{equation}
F=\frac{1}{2}(D\psi_1)(D\psi_1)^*+\frac{1}{2}(D\psi_2)(D\psi_2)^*
-\nu Re\Big\{(D\psi_1)(D\psi_2)^*\Big\}+\frac{1}{2}(\nabla\times {\bf A})^2 + F_p
\label{gl}
\end{equation}
Here
$D=\nabla + ie {\bf A}$, and $\psi_a=|\psi_a|e^{i\theta_a}$,
$a=1,2$, represent two superconducting  components  which,
 in a two-band superconductor
are associated with two different bands. Ref. \onlinecite{bcs}
discussed how coherence lengths are modified by interband
Josephson coupling. In Ref. \onlinecite{bcs2} the analysis was
carried out for a GL model with arbitrary interband interactions.
Here we briefly reproduce the key aspects
of the analysis from Refs. \onlinecite{bcs2, review}. In this analysis we  allow the term $F_p$ to
contain an {\it arbitrary} collection of non-gradient terms
representing various inter and intra-band interactions. I.e. $F_p$
includes but is not limited
 to interband Josephson coupling $\psi_1\psi_2^*+c.c.$  Below we
show how three characteristic length scales are defined in this two component model
(two are coherence lengths, associated with densities variations and
the London magnetic field penetration length), yielding type-1.5 regime $\xi_1 < \lambda < \xi_2$.
Note that
existence of two bands in a superconductor is {\it not} a sufficient conditions for
a superconductor to be described by a two-component Ginzburg-Landau model. Conditions of appearance of regimes
when the system does not allow
a description in terms of two-component fields
theory (\ref{gl}) is discussed in the work based on microscopic
considerations in Refs. \cite{silaev,silaev2}. However for a wide parameter 
range two-component GL expansions are justfied on formal grounds \cite{silaev2}

The effect of mixed gradient terms $-\nu Re\Big\{(D\psi_1)(D\psi_2)^*\Big\} $
 (which in this system
are induced by interband impurity scattering) was studied in
\cite{bcs2}. Below we consider the clean limit, thus we have to
set $\nu=0$, a reader interested in the effects of these term can
check the Ref \cite{bcs2}.  The Josephson coupling $\eta |\psi_1||\psi_2|\cos
(\theta_1-\theta_2)$ which is contained in $F_p$ tends to lock
phase difference  to $0$ or $\pi$ depending on sign of $\eta$. 
Lets denote the ground state values of
fields as $(|\psi_1|,|\psi_2|,\delta=(\theta_1-\theta_2))=(u_1,u_2,0)$ where $u_1>0$
and $u_2\geq 0$.
To
define coherence lengths one has to consider small deviations of the field
around ground state values ($\eps_1=|\psi_1|-u_1$,
$\eps_2=|\psi_2|-u_2$) and linearize the model in small deviations around the ground state (here we consider that
the phases are locked by Josephson coupling $\theta_1=\theta_2$)
\be F_{lin}=\frac12|\nabla\eps_1|^2+\frac12|\nabla\eps_2|^2+
\frac12\left(\begin{array}{c}\eps_1\\
\eps_2\end{array}\right)\cdot \hh\left(\begin{array}{c}\eps_1\\
\eps_2\end{array}\right)
\label{gllin}
+\frac12(\cd_1A_2-\cd_2A_1)^2
+\frac12e^2(u_1^2+u_2^2)|A|^2.
\ee
Here, $\hh$ is the Hessian matrix of $F_p(|\psi_1|,|\psi_2|,0)$
about $(u_1,u_2)$, that is,
\beq
\hh_{ab}=\left.\frac{\cd^2F_p}{\cd|\psi_a|\cd|\psi_b|}\right|_{(u_1,u_2,0)}.
\eeq
Now in $F_{lin}$, the vector potential $A$ decouples
and yields the London magnetic field penetration length
\beq
\lambda=\mu_A^{-1}=\frac{1}{e\sqrt{u_1^2+u_2^2}}
\eeq

In contrast, when there is interband coupling the density fields $\eps_1,\eps_2$ are, in general, coupled (i.e.\
in general the symmetric matrix $\hh$ has off-diagonal terms). To be able to define
coherence lengths that
coupling should be removed by a linear redefinition of fields \cite{bcs,bcs2,review},
\begin{equation}
 \chi_1=(|\psi_1|-u_1)\cos\Theta-(|\psi_2|-u_2)\sin\Theta\,, ~~~
\chi_2=-(|\psi_1|-u_1)\sin\Theta-(|\psi_2|-u_2)\cos\Theta\,.
\end{equation}
Then the linear theory decouples and thus allow to define two distinct coherence lengths
\be
F_{lin}=\frac12\sum_{a=1}^2\left(|\nabla\chi_a|^2+\mu_a^2\chi_a^2\right)
+\frac12(\cd_1A_2-\cd_2A_1)^2 +\frac12e(u_1^2+u_2^2)|A|^2. \ee
Thus the interband coupling, such as the Josephson terms
$\psi_1\psi_2^* +c.c.$ introduces hybridization of the fields. The
new coherence lengths are associated not with the fields $\psi_a$
but with their linear combinations $\chi_1,\chi_2$ \be
\xi_1=\mu_1^{-1}, \ee \be \xi_2=\mu_2^{-1} \ee

The type-1.5 regime occurs when $\xi_1<\lambda<\xi_2$, similarly
like in $U(1) \times U(1)$ theory. The difference with $U(1)\times U(1)$
theory is in temperature dependence of coherence lengths \cite{silaev,silaev2}.

\section{Vortex clusters in a Semi-Meissner state. Numerical results }

\begin{figure}[!htb]
  \hbox to \linewidth{ \hss
 \includegraphics[width=\linewidth]{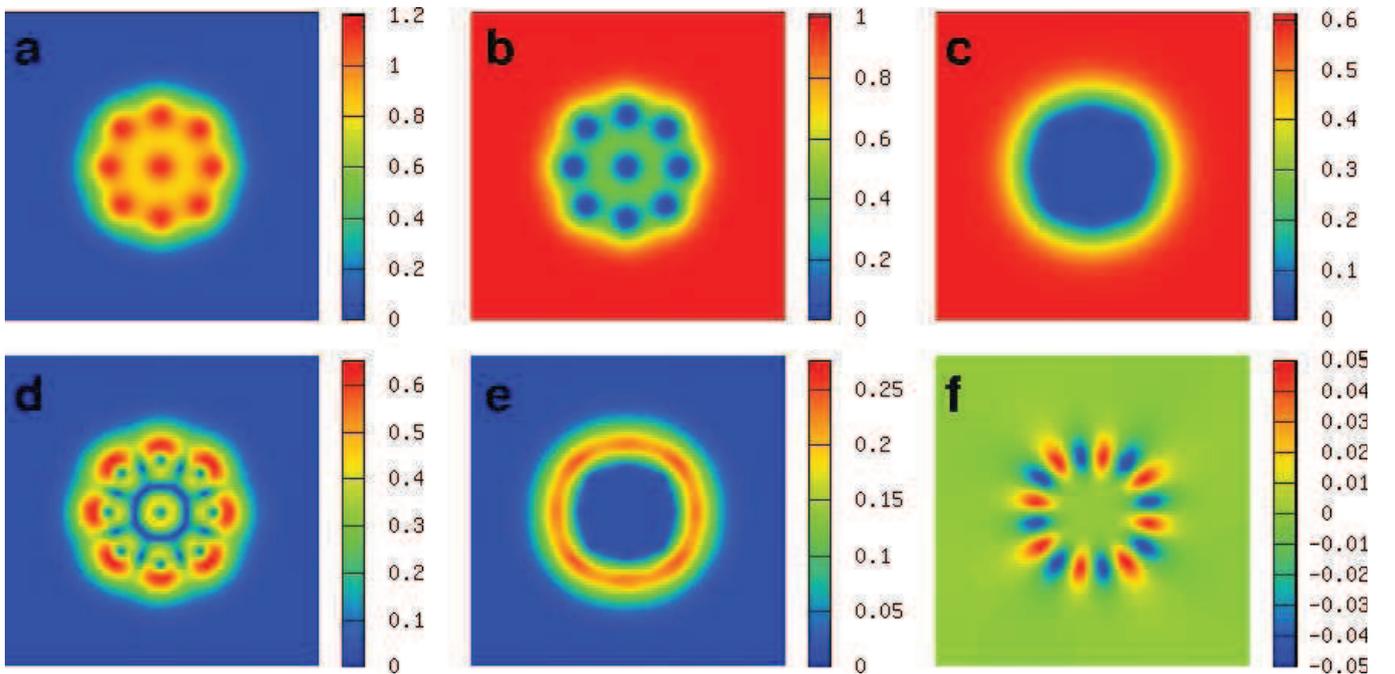}
\hss}
\caption{
Ground state of $N_v=9$ flux quanta in a type-1.5 superconductor enjoying $U(1)\times U(1)$
symmetry of the potential (\ie $\eta=0$) Credit: Ref. \cite{garaud}. The parameters of the potential being
here $(\alpha_1,\beta_1)=(-1.00, 1.00)$ and $(\alpha_2,\beta_2)=(-0.60, 1.00)$, while the electric
charge is $e=1.48$.
The displayed physical quantities are $\bf a$ the magnetic flux density,
$\bf b$ (\resp $\bf c$) is the density of the first (\resp second) condensate $|\psi_{1,2}|^2$.
$\bf d$ (\resp $\bf e$) shows the norm of the supercurrent in the first (\resp second) component.
Panel $\bf f$ is $\mathrm{Im}(\psi_1^*\psi_2)\equiv|\psi_1|| \psi_2|\sin(\theta_2-\theta_1)$ being nonzero
when there
appears a difference between the two condensates.
}
\label{2A-1}
\end{figure}

In this section, following Ref. \cite{garaud} we overview numerical solution for vortex
clusters in  two-component Ginzburg-Landau model
\begin{align}
\mathcal{F}&=
\frac{1}{2}\sum_{i=1,2}\Biggl[|(\nabla+ ie{\bf A}) \psi_i  |^2
+ (2\alpha_i+\beta_i|\psi_i|^2)|\psi_i|^2\Biggr]
+\frac{1}{2}(\nabla \times {\bf A})^2 -\eta|\psi_1|| \psi_2|\cos(\theta_2-\theta_1)
\label{FreeEnergy}
\end{align}
Here again,   $D=\nabla+ie{\bf A}$, and $\psi_i=|\psi_i| e^{i\theta_i}$.
Figure \ref{2A-1} (from Ref. \cite{garaud}) shows numerical solution for vortex cluster in type-1.5 superconductor with $U(1)\times U(1)$ symmetry.
The cases where $U(1)\times U(1)$ symmetry is weakly broken by interband Josephson coupling
are qualitatively similar \cite{garaud}.

From this numerical solution one can clearly see the hallmark
of type-1.5 superconductivity: the coexisting and
competing type-1 and type-2 behaviors of  the condensates despite
their electromagnetic coupling. Here the second component has a
type-1 like behavior, essentially its current is concentrated on
the surface of a vortex cluster similarly like current on a
surface of a type-1 superconductor, or on boundary
of macroscopic normal domains. On the contrary the  first
component tends to from well separated vortices.

\section{Discussion and Reply to Brandt and Das \cite{BrandtDas}.}

  In the proceedings of the previous conference of this series,
 E.H. Brandt and M. Das \cite{BrandtDas}
 criticized the idea of type-1.5 superconductivity. In this section we
discuss that these objections are  incorrect and  point out  that
 also unfortunately their overview of this idea and literature is highly inaccurate.
Also we discuss that  the following speculations
in \cite{BrandtDas} are unfortunately incorrect: namely
after dismissing the possibility of type-1.5 superconductivity in
two-band systems, the authors of \cite{BrandtDas} put forward a conjecture that
intervortex attraction reported by Moshchalkov et al \cite{moshchalkov}
might still be possible in $MgB_2$ but via some unrelated  
microscopic mechanisms like those which give intervortex
attraction in {\it single}-component systems with $\kappa \approx
1/\sqrt{2}$. As we point out below these mechanisms cannot lead to intervortex attraction in
superconductors like $MgB_2$ where the condition $\kappa \approx
1/\sqrt{2}$ does not hold.

First we would like to add some remarks
on the single-component $\kappa \approx 1/\sqrt{2}$ regime  in single-component systems
 and to mention some references on
pioneering works and  works of key importance, none of which  unfortunately  were cited  in  
Ref.\onlinecite{BrandtDas}. The idea of possible
non-monotonic intervortex potential in single-component models from microscopic corrections
(when the attractive and repusilve parts of forces at the GL-theory level  cancel each other out)   was first suggested by
Eilenberger and Buttner \cite{Eilenberger}, Halbritter \cite{Halbritter} and
  Dichtel \cite{Dichtel}.
 We stress that Brandt also should 
be credited for being one of the first who  followed up on these ideas, more precisely on the work by
Dichtel \cite{BrandtMF}.
 However, as discussed by Leung and Jacobs the mechanism
suggested by all the listed above early works, was unfortunately unrelated to the physics of
long-range vortex attraction in superconductors with $\kappa \approx 1/\sqrt{2}$ \cite{LeungJacobs}.
There was also a rather early works by
Jacobs \cite{Jacobs1} who was searching for intervortex attraction
in superconductors with $\kappa \approx 1/\sqrt{2}$ by generalized
the work of Neumann and Tewordt \cite{Neumann}.
The question that in weak-coupling superconductors with 
$\kappa \approx 1/\sqrt{2}$, microscopic corrections can lead to
intervortex attraction was put on
more solid grounds 
in the important work by Jacobs and Leung
\cite{LeungJacobs,Leung}. 

In the review of experiments on single component materials with
$\kappa \approx 1/\sqrt{2}$
 the Ref. \cite{BrandtDas}  reports two figures (Figs. 3 and 4 in \cite{BrandtDas})
 which are interpreted therein as experimental evidence for attractive
intervortex forces in Nb. However unfortunately these figures show 
evidence that actually an opposite effect takes place: namely they suggest that  long-range
intervortex forces are repulsive. Indeed the vortex clusters in
Fig. 4 form nearly hexagonal structure, which is only possible in
the systems with long-range repulsive interaction. Such vortex arrangements are
impossible if  long-range interaction were attractive. The Figs. 3
and 4 in Ref.\onlinecite{BrandtDas} show stripe and clump phases
which are known to occur in systems with two-scale purely
repulsive interactions \cite{structure1}. Indeed the
additional repulsive tail in the intervortex interaction can arise
because of  intervortex interaction via stray magnetic fields above the surface of a sample\cite{pearl}
(in fact the possibility that repulsive forces due to stray fields play role
in vortex structure formation in these experiments was already discussed by Kramer \cite{kramer}).
 Therefore the images
cited in \cite{BrandtDas} do not present  evidence for vortex
attraction in Nb, but rather present evidence to the contrary. A good review of current theoretical and
experimental evidence which is in favor of intervortex attraction
in Nb can be found in the  book by Huebener \cite{huebener}.

From a theoretical viewpoint the possibility of attractive
intervortex interaction for weak-coupling superconductors  with $\kappa \approx
1/\sqrt{2}$ is quite well established. However the conjecture that
this physics might somehow be relevant for systems like $MgB_2$ is
certainly incorrect. Namely, in these theories, in
single-component case one cannot get any appreciable intervortex
attraction through these mechanisms for $\kappa>1$. This is
because, this mechanism is  based on cancellation of forces at the
level of Ginzburg-Landau theory when $\xi \approx \sqrt{2}\lambda$. Introducing any, even small
coupling to a second band in general rapidly shrinks the parameter space
where a system can have  a regime analogous to Bogomolnyi limit ($\kappa
\approx 1/\sqrt{2}$). In case of multiple bands, achieving the
Bogomolnyi limit requires a fine tuning of GL parameters. 
That
is, in general multicomponent systems, in general, the region where GL interactions are
canceled out has zero measure  for all practical purposes. In
brief going to multicomponent system does not enhance the physics
of Bogomolnyi point ($\kappa \approx 1/\sqrt{2}$) but instead
dramatically shrinks the parameter space where this physics can occur. 

The
 Bogomolnyi  regime is indeed principally different from
  type-1.5 superconductivity. In type-1.5 regime the   non-monotonic interaction forces
is the consequence of the existence of several superconducting
components  with several coherence lengths such that $\xi_1<\sqrt{2}\lambda <\xi_2$.
The physics of type-1.5 regime is all about  the coexistence of components with type-1 and type-2 tendencies
 which is not possible in single component system.

Lets us also reply to the arguments, based on which 
the authors of Ref.\onlinecite{BrandtDas} dismissed the possibility
of type-1.5 superconductivity. They summarized their  argument as
{\sl In brief, mixing two components as visualized by Babaev et
al. and Moshchalkov et al., will not produce a new kind of
superconductivity, because in the mixture $\lambda$ (magnetic
field penetration length) is determined self-consistently.} Also
Ref.\onlinecite{BrandtDas} criticizes works on type-1.5
superconductivity by claiming that these works considered only
the limit of zero interband coupling between the superconducting
components from different bands. Ref.\onlinecite{BrandtDas}
also questioned whether  two coherence
length can be defined in multiband superconductors, namely: {\sl It is
shown that for the real superconductor which possesses a single
transition temperature, the assumption of two independent order
parameters with separate penetration depths and separate coherence
lengths is unphysical.}

Unfortunately all of these statements are factually not correct.
First, none of the papers on type-1.5 superconductivity ever
attributed different penetration lengths to different bands
(notations $\lambda_1,\lambda_2$ were used merely to parametrize 
Ginzbgurg-Landau model, as was very explicitly stated in \cite{bs1}).  Perhaps the origin of that misunderstanding
can be traced to a simple misconception of what constitutes intercomponent electromagnetic coupling,
reflected in the statements in
Sections 4 and 7   in  \onlinecite{BrandtDas}. Namely it is stated there
that the terms $\gamma_2 [D\psi_1(D\psi_2)^*
+c.c.]$ represent {\sl ``electromagnetic coupling between the
condensates"}. Unfortunately it is not correct because  these terms
are mixed gradient terms. The physical origin of these terms in two-band supercondctors is
multiband impurity scattering \cite{gurevich1,gurevich2} and thus
they    have little to
do with electromagnetic coupling. 
Moreover  even when such terms are present, in general they do not eliminate type-1.5 regime \cite{bcs2}.
In contrast the electromagnetic
coupling is entirely mediated by the vector potential $\bf A$ and
does not require any mixed gradient terms.
In
fact in  the Ref.\onlinecite{bs1} it was discussed in
considerable detail how penetration length is determined
self-consistently.

{Next the Ref. \onlinecite{BrandtDas}
 criticizes  the work
Ref.\onlinecite{bs1} for ``neglecting" interband Josephson coupling
and claims that  this coupling is ``generic". However first we should note that this
coupling is not generic and in some systems it is in fact forbidden on symmetry grounds
(corresponding references, on the systems where such situations occur were given in Ref.
\onlinecite{bs1,bcs2}). Second, we should remark that for the system where Josephson
coupling is present, its effects on type-1.5 regime was addressed
 \cite{bcs} long before the appearance of Ref.
\onlinecite{BrandtDas}, where this paper was not referenced. More recent papers studied effects of this
coupling in more detail \cite{bcs2,silaev,silaev2} and as reviewed
here indeed it does not eliminate the type-1.5 regime.

 The statement that in the presence of
interband coupling one cannot define two coherence lengths is a
more common misconception, also shared by other groups \cite{kogan}. 
For example the authors of Ref. \cite{kogan} came to such a conclusion 
after a failed attempt \cite{kogan1} to attribute different coherence lengths
directly to different gap fields neglecting hybridization. This naive 
approach is indeed  technically incorrect 
(see the Comment Ref. \onlinecite{bscomment}).
 As discussed above, two coherence
length are well defined when one takes into account hybridization:
i.e. when one attributes coherence lengths to different linear
combinations of the gap
fields\cite{bcs,bcs2,silaev,silaev2,review}. Also as reviewed
below two-component GL expansion is well justified on formal
grounds  under certain conditions even when interband coupling is present
\cite{silaev2}.

The statement about non-existence of two order parameters in the
abstract of Ref.\onlinecite{BrandtDas} is  indeed entirely irrelevant for
the existence  of type-1.5 regime. Note that
 in our works on $U(1)$ two-band systems
\cite{bcs,bcs2,silaev,silaev2,review}
were did not call  $\psi_{1,2}$  ``order parameters",
although this misnomer terminology is very commonly accepted recently. In two-band system $U(1)\times U(1)$ symmetry
is explicitly broken to $U(1)$ local symmetry.
Thus indeed the order parameter is a single complex field. However, 
in general, the number of components in the effective
 field theory (such as Ginzburg-Landau theory) has
 nothing to do with the notion of order parameters.
The number of components in GL theory is only related to question
whether or not the system is described by a multicomponent
classical effective field theory in some regime. One can have a perfectly
valid description of a system in terms of Ginzburg-Landau or
Gross-Pitaevskii complex fields theory even when there are no order parameters at
all and no spontaneously broken symmetries. The simplest examples are two-dimensional systems at finite temperature (where indeed $\psi$ cannot be called  an order parameter), a different example is  superfluid
turbulence. Similarly in two-band case the system can have
entirely well justified description in terms of {\it
two}-component GL theory with two distinct coherence lengths, despite having only $U(1)$ symmetry and
thus only a {\it  single} order parameter \cite{silaev2}.

To summarize this part: the   physics of
the $\kappa=1/\sqrt{2}$ regime has unfortunately no relationship to
type-1.5 superconductivity, which occurs in multicomponent systems
when $\xi_1<\sqrt{2}\lambda<\xi_2$. Also long-range vortex
attraction is necessary for type-1.5 regime but is not its
defining property (type-1.5 regime requires  at least two 
superconducting components). Finally two-component GL field theory and two
coherence lengths are well defined in case of interacting bands.
In the next section we review the recently published microscopic theory of the type-1.5
regime \cite{silaev,silaev2},  from which is it also quite apparent that this physics is principally different from the 
  physics of Bogomolnyi point ($\kappa \approx 1/\sqrt{2}$ in single-component theory).

\section{Microscopic theory of type-1.5 superconductivity}
\label{microscopic}

The phenomenological Ginzburg-Landau model described above
predicts the possibility of 1.5 superconducting state. This form
of two-component Ginzburg-Landau expansion was microscopically
justified on formal grounds \cite{silaev2}. Strictly speaking the
GL theory is justified only at elevated temperatures. To describe
type-1.5 superconductivity in all temperature regimes (except,
indeed the region where mean-field theory is inapplicable) as well
as to make a quantitative connection with a certain class of the real systems
requires a microscopic approach which also does not rely on a GL
expansion. The described above physics of type-1.5 regime was
recently justified by self-consistent Eilenberger theory
\cite{silaev}. We consider a superconductor with two overlapping
bands at the Fermi level \cite{suhl}. The corresponding two sheets
of the Fermi surface are assumed to be cylindrical. Within
quasi-classical approximation the band parameters characterizing
the two different sheets of the Fermi surface are the Fermi
velocities $v_{Fj}$ and the partial densities of states (DOS)
$\nu_j$, labelled by the band index $j=1,2$ and parameterized by
BCS pairing constants $\lambda_{11 (22)}$ (intraband) and
$\lambda_{12 (21)}$ (interband).

The asymptotic of the gap functions $\Delta_{1,2}(r)$ in two
superconducting bands at distances far from the vortex core can be
found by linearizing the Eilenberger equations together with the
self-consistency equations. The asymptotic of the linearized
system is governed by the complex plane singularities of response
function found in Ref.(\onlinecite{silaev}) among which are the
poles and branch cuts. {In general there are two regimes
regulating the asymptotic behavior of gap functions. The first
regime is realized when two poles of the response function lie
below the branch cut. The two poles determine the {\it two
coherence lengths} or, equivalently, the two masses of composite
gap functions fields (i.e. linear combinations of the fields as in
the previous section), which we denote as ``heavy" $\mu_H$ and
``light" $\mu_L$ (i.e. $\mu_H>\mu_L$). {\it This is principally
different from microscopic theories of the single-component
$\kappa \approx 1/\sqrt{2}$ regime}. At elevated temperatures
these masses (or inverse coherence lengths) are exactly the same
as the given by a corresponding two-component GL theory obtained
by the gradient expansion from the microscopic theory
\cite{silaev2}. The GL theory under certain conditions can be also
used 
at relatively low temperatures \cite{silaev2}.

The second regime is realized at lower temperatures when there is
only one pole of the response function lying below the branch cut
in the complex plane. In this case the asymptotic is determined by
the light mass mode $\mu_L$ and the contribution of the branch cut
which has all the length scales smaller than some threshold one
determined by the position of the lowest branch cut on the
imaginary axis. The branch cut contribution is essentially
non-local effect which is not captured by GL theory therefore one
can expect growing discrepancies between effective GL solution and
the result of microscopic theory at low temperatures.

The examples from Ref. \onlinecite{silaev} of the temperature
dependencies of the inverse coherence lengths $\mu_{L,H}(T)$  are
shown in the Fig.\ref{Fig:SequenceModes}. The evolution of the
inverse coherence lengths $\mu_{L,H}$ is shown in the sequence of
plots Fig.\ref{Fig:SequenceModes}(a)-(d) for $\lambda_J$
increasing from the small values $\lambda_J\ll
\lambda_{11},\lambda_{22}$ to the values comparable to intraband
coupling $\lambda_J\sim \lambda_{11},\lambda_{22}$. The two
massive modes coexist at the temperature interval $T^*_1<T<T_c$,
where the temperature $T^*_1$ is determined by the branch cut
position, shown in the Fig.\ref{Fig:SequenceModes} by black dashed
line. For temperatures $T<T^*_1$ there exists only one mode and as
the interband coupling parameter is increased, the temperature
$T^*_1$ rises and becomes equal to $T_c$ at some critical value of
$\lambda_J=\lambda_{Jc}$.

As shown on Fig.\ref{Fig:SequenceModes}a,b the function $\mu_L(T)$
is {\it non-monotonic} at low temperatures. The temperature
dependence of the inverse coherence length $\mu_L(T)$ has
anomalous behavior\cite{silaev}, which is in strong contrast to
temperature dependence of the mass of the gap mode in single-band
theories.

\begin{figure}
\centerline{\includegraphics[width=0.60\linewidth]{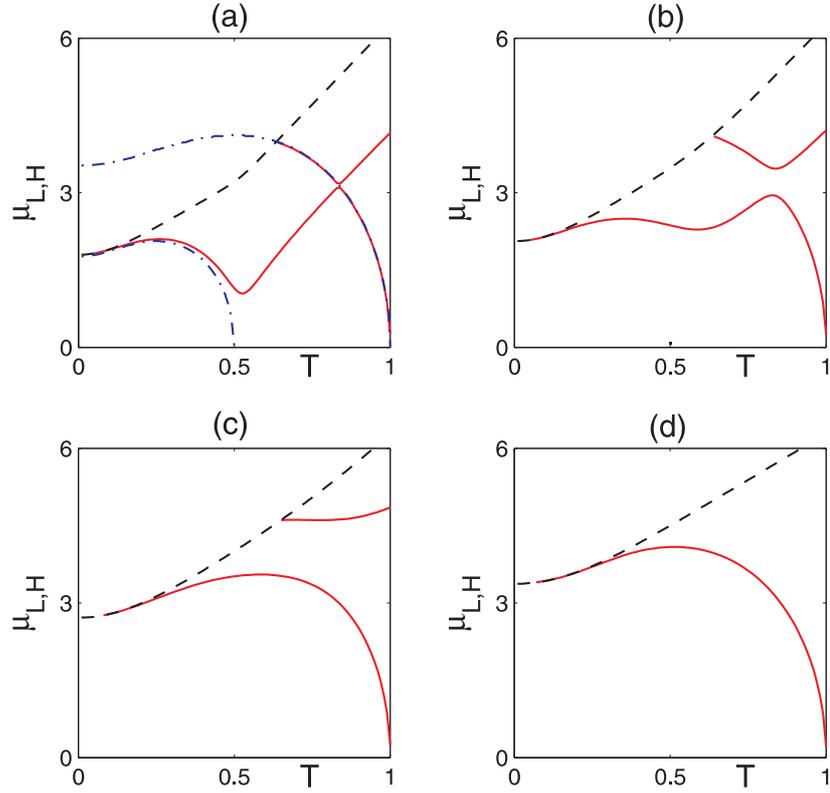}}
\caption{\label{Fig:SequenceModes} Calculated in Ref.
\onlinecite{silaev} two inverse coherence lengths $\xi_L^{-1}$ and $\xi_H^{-1}$.
Here the inverse coherence lenghth are the masses $\mu_{L}$ and $\mu_{H}$ (red solid
lines) of the composite gap function fields for the different
values of interband Josephson coupling $\lambda_J$ and
$\gamma_F=1$. In the sequence of plots (a)-(d) the transformation
of masses is shown for $\lambda_J$ decreasing from the small
values $\lambda_J\ll \lambda_{11},\lambda_{22}$ to the values
comparable to intraband coupling $\lambda_J\sim
\lambda_{11},\lambda_{22}$. The particular values of coupling
constants are $\lambda_{11}=0.25$, $\lambda_{22}=0.213$ and
$\lambda_J=0.0005;\;0.0025;\;0.025;\;\lambda_{22}$ for plots (a-d)
correspondingly. By black dash-dotted lines the branch cuts are
shown. In (a) with blue dash-dotted lines the masses of modes are
shown for the case of $\lambda_J=0$. Note that at $\lambda_J=0$
the two masses go to zero at two different temperatures. Because
$1/\mu_{L,H}$ are related to the coherence length, this reflects
the fact that for $U(1)\times U(1)$ theory there are two
independently diverging coherence lengths. Note that for finite
values of interband coupling only one mass $\mu_L$ goes to zero at
one $T_c$: this is in turn a consequence of the fact that
Josephson coupling breaks the symmetry down to single $U(1)$.  }
\end{figure}

To assess the effect of non-monotonic temperature dependence of
inverse coherence length $\mu_L(T)$ on the vortex structures in
two-band superconductors we calculated self-consistently
\cite{silaev,silaev2} the structure of isolated vortex for
different values of $\gamma_F=v_{F2}/v_{F1}$. {\it A complex aspect of
the vortex structure in two-band system is that in general the
exponential law of the asymptotic behavior of the gaps is {
not} directly related to the ``core size" at which gaps recover
most of their ground state values. } We can characterize this
effect by defining a ``healing" length $L_{\Delta i}$ of the gap
function as follows $|\Delta_i| (L_{\Delta i})= 0.95 \Delta_{i0}$.
The characteristic example of the vortex structure is shown in
Fig. \ref{Fig:VortexStructure15}a. For this case we obtain that
$L_{\Delta 1}\approx 0.8$ for all values of $\gamma_F$. On the
contrary, the healing length $L_{\Delta 2}$ of changes
significantly such that $L_{\Delta 2}= 1.6;\; 2.5;\; 3.2;\;
3.9;\;4.5$ for $\gamma_F=1;\; 2;\; 3;\; 4;\; 5$ correspondingly.

\begin{figure}
\centerline{\includegraphics[width=0.60\linewidth]{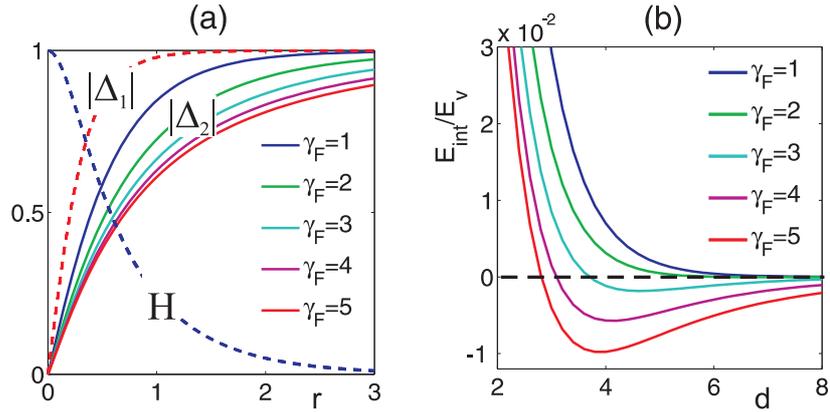}}
\caption{\label{Fig:VortexStructure15} Calculated in Ref.
\onlinecite{silaev} (a) Distributions of magnetic field
$H(r)/H(r=0)$, gap functions $|\Delta_1|(r)/\Delta_{10}$ (dashed
lines) and $|\Delta_2|(r)/\Delta_{20}$ (solid lines) for the
coupling parameters $\lambda_{11}=0.25$, $\lambda_{22}=0.213$ and
$\lambda_{21}=0.0025$ and different values of the band parameter
$\gamma_F=1;2;3;4;5$. (b) The energy of interaction between two
vortices normalized to the single vortex energy as function of the
intervortex distance $d$.  It clearly exhibits long-range attraction, short-range
repulsion as a consequence of two coherence lengths in the type-1.5 regime.{ In panels (c,d) the temperature is
$T=0.6$.}
 }
\end{figure}

The temperature dependencies of the sizes of the vortex cores in
two superconducting bands calculated in \cite{silaev,silaev2} in
the full nonlinear model according to the two alternative
definitions. The first one is the slope of the gap function
distribution at $r=0$ which characterizes the width of the vortex
core near the center $R_{cj}=(d\ln\Delta_{j}/dr)^{-1}(r=0)$
[Fig.(\ref{Fig:CoreSize})a]. The second one is the healing length
$L_{hj}$ defined as $\Delta_{j}(L_{hj})=0.95\Delta_{0j}$
[Fig.(\ref{Fig:CoreSize})b] (i.e. this length is not directly
related to exponents but quantifies at what length scales the gap
functions almost recover their ground state values). Both
definitions demonstrate the stretching of the vortex core in the
weak component related to the peak of the coherence length shown
in the Fig.(\ref{Fig:SequenceModes})a. Note that the weak band
healing length $L_{h2}(T)$ in Fig.(\ref{Fig:CoreSize})b has
maximum at the temperature slightly larger than $T_{c2}$ which is
consistent with the fact that the maximum of coherence length
$\xi_L$ (equivalently the minimum of the field mass $\mu_L$) in
Fig.\ref{Fig:SequenceModes}a is shifted to the temperature above
$T_{c2}$  ($T_{c2}$ is defined as the lower critical temperature
in the limit of no Josephson coupling).

Besides justifying the predictions of phenomenological
two-component GL theory \cite{silaev2} the microscopic formalism
developed in Ref.\onlinecite{silaev} allows to describe type-1.5
superconductivity beyond the validity of GL models. The type-1.5
behavior requires a density mode with low mass $\mu_L$ to mediate
intervortex attraction at large separations, which should coexist
with short-range repulsion. The non-monotonic temperature behavior
of the inverse coherence length $\mu_L(T)$ shown in
Fig.(\ref{Fig:SequenceModes})a,b makes possible the attractive
interaction between vortices. Furthermore  because the softest
mode with the mass $\mu_L$ in two band system may be associated
with only a fraction of the total condensate (as follows from
corresponding mixing angles), and because there could be the
second mixed gap mode with larger mass $\mu_H$, the short-range
intervortex interaction can be repulsive marking the transition to
the type 1.5 regime at low temperatures.

\begin{figure}[h!]
\includegraphics[width=0.60\linewidth]{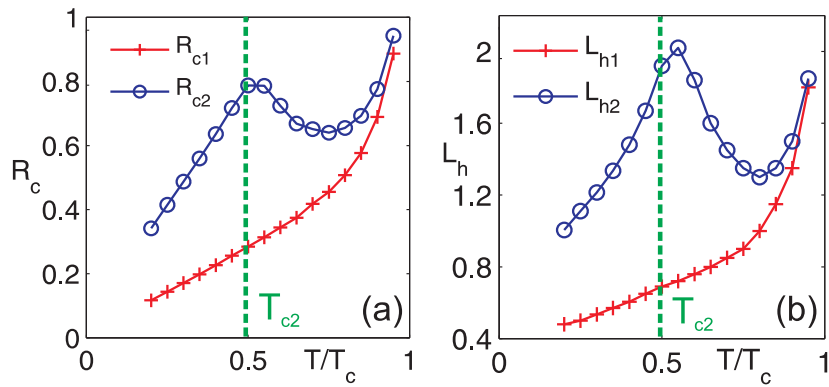}
\caption{\label{Fig:CoreSize} Calculated in Ref.
\onlinecite{silaev2} (a) Sizes of the vortex cores $R_{c1,2}$ and
(b) healing lengths $L_{h1,2}$ in weak (blue curve, open circles)
and strong bands (red curve, crosses) as functions of temperature.
The parameters are $\lambda_{11}=0.5$, $\lambda_{22}=0.426$,
$\lambda_{12}=\lambda_{21}=0.0025$ and $v_{F2}/v_{F1}=1$. In the
low temperature domain, the vortex core size in the weak component
grows and reaches a local maximum near the temperature $T_{c2}$
(the temperature near which the weaker band crosses over from
being active to having superconductivity induced by an interband
proximity effect) \cite{silaev}. In the absence of interband
coupling there is a genuine second superconducting phase
transition at $T_{c2}  = 0.5 T_{c1}$ where the size of the second
core diverges. When interband coupling is present  it gives an
upper bound to the core size in this temperature domain,
nonetheless this regime is especially favorable for appearance of
type-1.5 superconductivity \cite{silaev,silaev2}. }
\end{figure}

 The microscopically demonstrated
existence of two well defined coherence lengths
\cite{silaev,silaev2} and disparity of the characteristic length
scales of variations of the densities two superconducting
components shown in Fig.(\ref{Fig:VortexStructure15}a) and
(\ref{Fig:CoreSize}) results in the type-1.5 superconductivity
with physical consequences summarized in the table \ref{table1}.

\section{Conclusion}
We reviewed the concept of type-1.5 superconductivity in multicomponent systems.
Both at the levels of microscopic and  Ginzburg-Landau   theories
the behavior arises as a consequence of the existence
of several superconducting components with different coherence lengths  $\xi_{1,2}$  in the system.
In the type-1.5 regime one or several of coherence lengths $\xi_{1,2}$ are larger than the magnetic field
penetration length $\lambda$,  while other coherence lengths are smaller than $\lambda$.
These coherence lengths are well
defined not only for $U(1)\times U(1)$ systems but also (under certain conditions) in case
of multi-band systems with only $U(1)$ symmetry. The concept also
arises in systems with larger number of components and in particular
in three band systems with broken time reversal, where the
broken symmetry is $U(1)\times Z_2$ \cite{3bands}.
The properties
of this state are summarized in table \ref{table1}. A more detailed review of this state
is available in \cite{review}.

\section{Acknowledgments}
EB was supported by Knut and Alice Wallenberg
Foundation through the Royal Swedish Academy of Sciences, Swedish Research Council and by the US National
Science Foundation CAREER Award No. DMR-0955902.
 MS was  supported by the Swedish
Research Council, "Dynasty" foundation, Presidential RSS Council
(Grant No. MK-4211.2011.2) and Russian Foundation for Basic
Research.
The computations were performed on resources
provided by the Swedish National Infrastructure for Computing (SNIC)
 at National Supercomputer Center at Linkoping, Sweden.


\end{document}